\documentstyle[12pt]{article}
\newlength{\extraspace}
\newlength{\extraspaces}
\newcommand{\be}{\begin{equation}}
\newcommand{\ee}{\end{equation}}
\newcommand{\ba}{\begin{eqnarray}}
\newcommand{\ea}{\end{eqnarray}}

\newcommand{\nonu}{\nonumber \\[.5mm]}
\newcommand{\A}{&\!\!\!}

\begin{document}

\thispagestyle{empty}
\begin{flushright}
SIT-LP-06/11 \\
{\tt hep-th/0611072} \\
November, 2006
\end{flushright}
\vspace{7mm}
%
\begin{center}
{\large \bf  On Origin of Mass and Supersymmetry }  
\\[20mm]
{\sc Kazunari Shima}
\footnote{
\tt e-mail: shima@sit.ac.jp}
\ and \
{\sc Motomu Tsuda}
\footnote{
\tt e-mail: tsuda@sit.ac.jp}
\\[5mm]
${}^{\rm a}${\it Laboratory of Physics, 
Saitama Institute of Technology \\
Fukaya, Saitama 369-0293, Japan} \\[20mm]
%
%

\begin{abstract}
We show that the nonlinear supersymmetric general relativity 
gives new insights into the origin of mass and elucidates the mysterious relations 
between the cosmology and the (low energy) particle physics. 
\end{abstract}
\end{center}

\newpage

\noindent
In the previous papers we have shown that Nonlinear Supersymmetric General Relativity (NLSUSY GR) 
\cite{s1} may give a new insight into the unsolved problems of the cosmology, e.g. 
the dark matter, the dark enery and the mysterious relations between the cosmology and  
the (low energy) particle physics, etc. \cite{st1}. 
In this article we study further these problems in self-contained and explicit arguments. 

NLSUSY GR is obtained by extending the geometrical arguments of Einstein general relativity (EGR) 
to new (called {\it SGM} from composite viewpoints \cite{s1}) spacetime 
just inspired by NLSUSY, where {\it tangent spacetime} is specified not only 
by the ordinary SO(3,1) Minkowski coordinates $x^{a}$ but also 
by the SL(2,C) Grassmann coordinates $\psi$ 
turning subsequently to the dynamical Nambu-Goldstone (NG) fermion (called {\it superon} hereafter) 
of NLSUSY Volkov-Akulov (VA) model \cite{va}. NLSUSY GR ($N = 1$ SGM) action \cite{s1} is written 
as the following geometrical form of (empty) SGM spacetime (of everything): 
\begin{equation}
L(w)={c^{4} \over 16{\pi}G}\vert w \vert(\Omega(w) - \Lambda), 
\label{SGM}
\end{equation}
where 
\begin{equation}
\vert w \vert=\det{w^{a}}_{\mu}=\det({e^{a}}_{\mu}+ {t^{a}}_{\mu}(\psi)),  \quad
{t^{a}}_{\mu}(\psi)={\kappa^{2}  \over 2i}(\bar{\psi}\gamma^{a}
\partial_{\mu}{\psi}
- \partial_{\mu}{\bar{\psi}}\gamma^{a}{\psi}). 
\label{w}
\end{equation} 
$w^{a}{_\mu}(x)$, $\Omega(w)$, $s_{\mu\nu} \equiv w^{a}{_\mu}\eta_{ab}w^{b}{_\nu}$ and 
$s^{\mu \nu}(x) \equiv w^{\mu}{_{a}}(x) w^{{\nu}{a}}(x)$ 
are the fundamental unified quantities defining SGM spacetime \cite{st2,st3}, 
i.e. the (composite) vierbein, 
the scalar curvature and the unified metric tensors of SGM spacetime. 
${e^{a}}_{\mu}$ for the local SO(3,1) and 
${t^{a}}_{\mu}(\psi)$ for the local SL(2,C) are the ordinary vierbein of EGR 
and the mimic vierbein of the stress-energy-momentum tensor 
of superon $\psi(x)$, whose dynamical parallel alignments for the spacetime orientations of 
SGM spacetime are encoded in $w^{a}{_\mu}(x)$.
$G$ is the Newton gravitational constant, $\Lambda$ is a ({\it small}) cosmological term.  
Remarkablly ${\kappa}$, which is the arbitrary constant of NLSUSY VA model with the dimension $(length)^2$ 
and subsequently becomes the strength of the superon-vacuum coupling constant in the (low energy theorem) 
particle physics viewpoint, is now related to $G$ and $\Lambda$ 
by ${\kappa^{2} = ({c^{4}\Lambda \over 8{\pi}G}})^{-1}$ 
in the NLSUSY GR (SGM) scenario. 
NLSUSY GR action (\ref{SGM}) is invariant at least under 
$[{\rm new \ NLSUSY}] \otimes [{\rm local \ GL(4,R)}] 
\otimes [{\rm local \ Lorentz}]$ $\otimes \ [{\rm local \ spinor \ translation}]$ 
for spacetime symmetries 
%
and 
[global SO($N$)] $\otimes$ [local ${\rm U}(1)^N$] 
for the internal symmetries 
in $N$(-superons $\psi^i$ ($i = 1, 2, \cdots, N$))-extended NLSUSY GR \cite{st2}. 
%

NLSUSY GR is unstable due to the NLSUSY structure of tangent space 
and decays spontaneously (called {\it Big Decay} of spacetime) to ordinary Riemann spacetime 
described by the Einstein-Hilbert action 
coupled with massless superon (matter) ${\psi}$ of NLSUSY VA model, 
which is written formally as follows, 
\begin{equation}
L(e, \psi)={c^{4} \over 16{\pi}G}\vert e \vert \{ R(e) - \Lambda + T(e, \psi) \}, 
\label{SGMR}
\end{equation}
where $T(e, \psi)$ is the kinetic term and the gravitational interaction of superon. 
The second and the third terms produce ($N$-extended) NLSUSY VA action 
in asymptotic Riemann-flat ($e{^a}_{\mu}(x) \rightarrow {\delta}{^a}_{\mu}$) spacetime 
with ${\kappa^{2} = ({c^{4}\Lambda \over 8{\pi}G}})^{-1}$, 
which is different from SGM-flat ($w{^a}_{\mu}(x) \rightarrow {\delta}{^a}_{\mu}$) spacetime for (\ref{SGM}). 

As some NL theories can be recasted into equivalent linear (L) renormalizable theories,  
NLSUSY is also the case and the equivalence of $N = 1$ VA model to the free action 
of $N = 1$ scalar mutiplet or of $N = 1$ axial vector multiplet are studied in detail \cite{ikruz}. 
The linearization of the highly nonlinear SGM action (\ref{SGMR}) 
may be inevitable to extract the physical contents of SGM, 
where the broken LSUSY is realized on the supermultiplet containing mass gaps. 
To investigate the mass (gap) generation mechanism in SGM scenario  
it is useful to go to asymptotic flat spacetime. 
We have shown \cite{stt1} that (the more realistic) $N = 2$ NLSUSY VA model 
in terms of two superons (Majorana NG fermion fields) $\psi^i$ ($i = 1, 2$), 
which corresponds to the SGM action (\ref{SGMR}) for $N = 2$ in asymptotic Riemann-flat spacetime, 
is (algebraically) equivalent to the spontaneouly broken $N = 2$ LSUSY massless supermultiplet 
with component fields denoted by $(v_a, \lambda^i, \phi, F^I)$ $(I = 1, 2, 3)$ 
for a U(1) gauge field, 
two Majorana spinor fields of $\bf{\underline {2}}$  of SU(2), 
a complex scalar field 
and three auxiliary real scalar fields of $\bf{\underline {3}}$ of SU(2); 
indeed, the $N = 2$ NLSUSY VA action, 
\begin{equation}
L_{\rm VA}  =  - {1 \over {2 \kappa^2}} \vert w \vert 
 =  - {1 \over {2 \kappa^2}} \left\{ 1 + t{^a}_a 
+ {1 \over 2} (t{^a}_a t{^b}_b - t{^a}_b t{^b}_a) 
+ \cdots \right\} 
\label{vaactex}
\end{equation}
with $\vert w \vert = \det w^a{}_b = \det (\delta^a_b + t^a{}_b)$  
and $t^a{}_b = - i \kappa^2 \bar\psi^i \gamma^a \partial_b \psi^i$, 
is recasted into the equivalent spontaneouly broken $N = 2$ LSUSY action 
\begin{equation}
L_0 =  \partial_a \phi \partial^a \phi^* 
- {1 \over 4} F^2_{ab} 
+ {i \over 2} \bar\lambda^i \!\!\not\!\partial \lambda^i 
+ {1 \over 2} (F^I)^2 - {1 \over \kappa} \xi^I F^I, 
\label{lact}
\end{equation}
where $F_{ab} = \partial_a v_b - \partial_b v_a$, 
the $\xi^I$ are arbitrary real parameters for the induced global 
SO(3) (SU(2)) rotation satisfying $(\xi^I)^2 = 1$ 
and $L_0$ means the (almost) free action. 
As usual in the linearization of NLSUSY the Fayet-Iliopoulos (FI) $F$ terms in (\ref{lact}) 
indicating spontaneous SUSY breaking with the vacuum expectation value 
$<F^I> = {\xi^I \over \kappa}$ are generated {\it automatically}. 
And all component fields of the LSUSY supermultiplet can be constructed 
as the composites of superons $\psi^i$ (called SUSY invariant relations \cite{ikruz}),  
%
in such a way as the familiar LSUSY transformations \cite{wb} 
defined on the elementary supermultiplet are reproduced on the abovementiond composite supermultiplet 
in terms of the NLSUSY transformations, 
$\delta_\zeta \psi^i = {1 \over \kappa} \zeta^i 
- i \kappa \bar\zeta^j \gamma^a \psi^j \partial_a \psi^i$, 
on the constituent superons $\psi^i$ \cite{stt1}. 
For $\xi^1 = \xi^2 = 0$ the above supermultiplet 
is the ordinary $N = 1$ (axial) vector multiplet containing the U(1) gauge field. 

We have tried to extract the cosmological implications of SGM scenario 
described by (\ref{SGM}) and the subsequent Big Decay to Riemann spacetime with massless superon (\ref{SGMR}) 
igniting the Big Bang \cite{st3}. 
The variation of (\ref{SGMR}) with respect to  ${e^{a}}_{\mu}$ gives 
the equation of motion for ${e^{a}}_{\mu}$ depicted formally as follows 
\begin{equation}
R_{\mu\nu}(e)-{1 \over 2}g_{\mu\nu}R(e) = 
{8{\pi}G \over c^{4}} 
\left\{ T_{\mu\nu}(e,{\psi})-g_{\mu\nu}{c^{4}\Lambda \over 8{\pi}G} \right\}, 
\label{SGMEQ}
\end{equation}
where $T_{\mu\nu}(e,{\psi})$ represents the stress-energy-momentum of superon  
including the gravitational interactions. 
Note that $-{c^{4}\Lambda \over 8{\pi}G}$ can be interpreted as 
{\it the constant negative energy density of empty spacetime}, 
i.e. {\it the dark energy density ${-{\rho}_{D}}$} responsible for the 
observed present accelleration of the universe. 
%
%
While, on tangent spacetime, the low energy theorem of the particle physics 
gives the following superon-vacuum coupling 
${<{\psi^j}_{\alpha}(q) \vert {J^{k\mu}}_{\beta} \vert 0> = 
i\sqrt{c^{4}\Lambda \over 8{\pi}G}(\gamma^{\mu})_{\alpha\beta} \delta^{jk} e^{iqx}+ \cdots}$,  
where ${{J^{k\mu}}= i\sqrt{c^{4}\Lambda \over 8{\pi}G} \gamma^{\mu}\psi^k + \cdots}$      
is the conserved supercurrent of NLSUSY VA action \cite{s2}        
and $\sqrt{c^{4}\Lambda \over 8{\pi}G}$ is {\it the coupling constant $g_{sv}$ 
of superon with the vacuum}. 
Further as shown before the right hand side of (\ref{SGMEQ}) for $N = 2$ flat case 
is essentially $N = 2$ NLSUSY VA action. 
And it is equivalent to the broken LSUSY action (\ref{lact}) which contains 
the SUSY breaking mass (gap) $M_{SUSY}$ 
for the (composite) fields of the (massless) LSUSY supermultiplet 
\begin{equation}
{M_{SUSY}}^{2} \sim <F^{I}> \sim {\sqrt{c^{4}\Lambda \over 8{\pi}G}}\xi^{I}  \sim g_{sv}\xi^{I}.  
\label{relation}
\end{equation}
Suppose that among the LSUSY supermultiplet 
the stable and the lightest particle retains the mass of the order of the spontaneous SUSY breaking ${M_{SUSY}}$. 
And if the neutrino is such a particle $\lambda(x)$, 
i.e. 
${m_{\nu}}^{2} \sim \sqrt{c^{4}\Lambda \over 8{\pi}G}$, 
then SGM predicts the observed value of the (dark) energy density of the universe and 
naturally explains the mysterious coincidence between $m_{\nu}$ and $\rho^{obs} \sim \rho_{D}$,
\begin{equation}
\rho^{obs} \sim (10^{-12}GeV)^{4} \sim {m_{\nu}}^{4}.
\label{DENSITY}
\end{equation}
The tiny values of neutrino mass and of the energy density of the universe are 
the direct evidence of SUSY breaking 
in the spontaneous phase transition of SGM spacetime (Big Dicay) preceding the Big Bang. 

In the above discussions the realistic mass generation mechanism in the LSUSY supermultiplet is speculative, 
for unfortunately the ordinary Yukawa interaction term is absent in the equivalent LSUSY theory (\ref{lact}). 
Recently we have found that the ordinary LSUSY invariant mass terms 
as well as the Yukawa interaction terms vanish identically under the SUSY invariant relation 
in the linearization of NLSUSY. 
Therefore we can add these terms to the linearized free action (\ref{lact}) 
{\it without violating the equivalence} and the invariance \cite{st4}. 
Now we can investigate explicitly the realistic mass generation mechanism in SGM scenario. 

For simplicity and without loss of generality 
we consider $N = 2$, $D = 2$ SO(2) SUSY and $D = 2$ SGM. 
The NLSUSY VA model has the same form as (\ref{vaactex}) 
except $L_{\rm VA}$ terminates at ${\cal O}(t^2)$. 
The equivalent $D = 2$ LSUSY theory \cite{st4} consists of 
the supermultiplet with component fields $(v^a, \lambda^i, A, \phi, F)$ 
and the following Lagrangian  
\be
L_0 = {1 \over 2} (\partial_a A)^2 
- {1 \over 4} (F_{ab})^2 
+ {i \over 2} \bar\lambda^i \!\!\not\!\partial \lambda^i 
+ {1 \over 2} (\partial_a \phi)^2 
+ {1 \over 2} F^2 
- {1 \over \kappa} \xi F. 
\label{LSUSYact}
\ee
We constructed the action (\ref{LSUSYact}) 
based on the arguments in $N = 1$, $D = 2$ super-Yang-Mills theory. 
The $D = 2$ SUSY invariant relations for those component fields 
are constructed as composites of $\psi^i$ in all orders as follows \cite{st4}, 
\ba
\A \A 
v^a = - {i \over 2} \xi \kappa \epsilon^{ij} \bar\psi^i \gamma^a \psi^j \vert w \vert, 
\label{expand-v}
\\
\A \A 
\lambda^i = \xi \left[ \psi^i \vert w \vert 
- {i \over 2} \kappa^2 \partial_a 
\{ \gamma^a \psi^i \bar\psi^j \psi^j 
(1 - i \kappa^2 \bar\psi^k \!\!\not\!\partial \psi^k) \} \right], 
\\
\A \A 
A = {1 \over 2} \xi \kappa \bar\psi^i \psi^i \vert w \vert, 
\\
\A \A 
\phi = - {1 \over 2} \xi \kappa \epsilon^{ij} \bar\psi^i \gamma_5 \psi^j 
\vert w \vert, 
\\
\A \A 
F = {\xi \over \kappa} \vert w \vert 
- {1 \over 8} \xi \kappa^3 
\Box ( \bar\psi^i \psi^i \bar\psi^j \psi^j ). 
\label{expand-F}
\ea 
The most general LSUSY invariant (Yukawa) interaction  and mass terms are given by 
\ba
L_f = f ( A \bar\lambda^i \lambda^i - \epsilon^{ij} \phi \bar\lambda^i \gamma_5 \lambda^j+ A^2 F - \phi^2 F - \epsilon^{ab} A \phi F_{ab} )  
\label{yukawa}
\ea
and 
\ba
L_m = - {1 \over 2} m \ ( \bar\lambda^i \lambda^i - 2 A F + \epsilon^{ab} \phi F_{ab}),  
\label{mass}
\ea
where $f$ is an arbitrary constant with the dimension $(mass)^1$. 
Substituting the SUSY invariant relations from (\ref{expand-v}) to (\ref{expand-F}) into (\ref{yukawa}) and (\ref{mass}) 
we see they vanish identically, i.e. 
$L_{f}(\psi)=L_{m}(\psi) \equiv 0$ in $D = 2$ (and $D = 4$ as well). 
Therefore the most general form of the renormalizable LSUSY invariant theory equivalent 
to $N = 2$, $D = 2$ NLSUSY VA action is written as follows, 
\ba
L_{\rm VA}=L_{0} + L_{f} + L_{m}, 
\label{equiL}
\ea
which is equipped with the self-contained spontaneous SUSY breaking. 

Now at the tree level we can easily see the mediation of mass genaration 
in the spontaneous SUSY breaking 
from the spacetime origin for NLSUSY to the (Yukawa-)Higgs origin for LSUSY as follows. 
For simplicity and correponding to the chiral LSUSY supermultiplet we put $L_{m} = 0$.
The configuration of the scalar fields for the vacuum determines the contents of the LSUSY theory. 
The  potential is given by  
\begin{equation}
V(A, \phi, F)= - {1 \over 2} F^2  + {\xi \over \kappa} F - f (A^2 F - \phi^2 F). 
\label{pot1}
\end{equation}
Substituting the field equation for the auxiliary field, 
$F={\xi \over \kappa}-f(A^{2}-\phi^{2})$, into (\ref{pot1}) 
we obtain 
\begin{equation}
V(A, \phi) = {1 \over 2} 
\left\{ f(A^2-\phi^2)-{\xi \over \kappa} \right\}^2 \geq 0. 
\label{pot2}
\end{equation}
In the $(A,\phi)$ plane, the massless mode and the massive mode appear parallel 
and perpendicular to the hyperbolic curve, 
$A^{2}-\phi^{2}={\xi \over f\kappa}$, respectively and some staffs of the LSUSY supermultiplet become massive. 
To see this we express the scalar fields A and $\phi$ as follows, 
\begin{equation}
A(x)=-\ \sqrt{{\xi \over {f\kappa}}\{1+\varphi(x)\}}\cosh{\theta(x)}
\label{A}
\end{equation}
\begin{equation}
\phi(x)=\sqrt{{\xi \over {f\kappa}}\{1+\varphi(x)\}}\sinh\theta(x) 
\label{phi}
\end{equation}
and substitute these expressions into $L_{0}+L_{f}$. 
We obtain after rescaling appropriately 
\ba
L_{0}+L_{f} \A = \A 
{1 \over 2}(\partial_{a}\varphi)^{2} 
+ {1 \over 2}(\partial_{a}\theta)^{2} - {1 \over 4} F^2_{ab} 
+ {i \over 2} \bar\lambda^i \!\!\not\!\partial \lambda^i 
- \sqrt{f\xi \over \kappa}\bar\lambda^i \lambda^i
-{2f\xi \over \kappa}{\varphi}^{2} 
\nonu
\A \A 
+ \cdots ({\rm interaction \ terms}), 
\label{linmass}
\ea
where we assume $\varphi, \theta \ll 1$ for simplicity. 
We see from Eq.(\ref{linmass}) that the fermions $\lambda^i$ and the radial scalar mode $\varphi$ 
get the following masses in $D = 2$ SGM scenario 
with ${\kappa^{2} = ({c^{4}\Lambda \over 8{\pi}G}})^{-1}$, 
\begin{equation}
m_{\lambda^i}^{2} = \ m_{\varphi}^{2} 
= 4\ \sqrt{c^{4}\Lambda \over 8{\pi}G}{f\xi},   
\label{masses}
\end{equation}
while $\theta$ and $v_{a}$ remain massless. 
These arguments show that the relation (\ref{relation}) and (\ref{DENSITY}) 
hold in $D = 2$ and $D = 4$ as well, provided $f\xi \sim {\cal O}(1)$. 
(The same order contribution of the see-saw mechanism are anticipated \cite{numass}.) 
%
%
%
These results show explicitly the mediation of mass genaration 
in the spontaneous SUSY breaking in SGM scenario 
from the spacetime origin in NLSUSY to (Yukawa-)Higgs origin in LSUSY. 
We can perform the same arguments for $D = 4$ $N$-extended SUSY 
by using the chiral representations (for fermions), 
where the induced grobal SU($N$) internal symmetry 
are formulated manifestly and compactly as shown for $D = 4$ $N = 2$ \cite{stt1}. 
The large mass scales and the compact (broken) gauge d.o.f. necessary 
for constructing the realistic model 
will appear through the linearization of $T_{\mu\nu}(e,{\psi})$ with the large $N$ 
which contains the mass scale $\Lambda^{-1}$ and the higher order terms with $\psi$. 
NLSUSY GR with extra spacetime dimensions equipped 
with the Big Decay and the spontaneous compactification 
is interesting and in the same scope.

\newpage

%
\newcommand{\NP}[1]{{\it Nucl.\ Phys.\ }{\bf #1}}
\newcommand{\PL}[1]{{\it Phys.\ Lett.\ }{\bf #1}}
\newcommand{\CMP}[1]{{\it Commun.\ Math.\ Phys.\ }{\bf #1}}
\newcommand{\MPL}[1]{{\it Mod.\ Phys.\ Lett.\ }{\bf #1}}
\newcommand{\IJMP}[1]{{\it Int.\ J. Mod.\ Phys.\ }{\bf #1}}
\newcommand{\PR}[1]{{\it Phys.\ Rev.\ }{\bf #1}}
\newcommand{\PRL}[1]{{\it Phys.\ Rev.\ Lett.\ }{\bf #1}}
\newcommand{\PTP}[1]{{\it Prog.\ Theor.\ Phys.\ }{\bf #1}}
\newcommand{\PTPS}[1]{{\it Prog.\ Theor.\ Phys.\ Suppl.\ }{\bf #1}}
\newcommand{\AP}[1]{{\it Ann.\ Phys.\ }{\bf #1}}


\begin{thebibliography}{100}
%
\bibitem{s1} K. Shima, {\it Phys. Lett.} {\bf B501} (2001) 237. \\
%
             K. Shima, {\it European Phys. J.} {\bf C7} (1999) 341. 
%
\bibitem{st1} K. Shima and M. Tsuda, {\it PoS HEP2005} (2006) 011. 
%
\bibitem{va}  D.V. Volkov and V.P. Akulov, {\it Phys. Lett.} {\bf B46} (1973) 109. 
%
\bibitem{st2} K. Shima and M. Tsuda, {\it Phys. Lett.} {\bf B507} (2001) 260. 
%
\bibitem{st3} K. Shima and M. Tsuda, {\it Class. and Quantum. Grav.} {\bf 19} (2002) 5101. \\
%
              K. Shima, M. Tsuda and M. Sawaguchi, {\it Int. J. Mod. Phys.} {\bf E13} (2004) 539. 
\bibitem{ikruz} E.A. Ivanov and A.A. Kapustnikov, {\it J. Phys.\ }{\bf A11} (1978) 2375. \\
%
             M. Ro\v{c}ek, \PRL{41} (1978) 451. \\
%
             T. Uematsu and C.K. Zachos, \NP{B201} (1982) 250. \\ 
%
             K. Shima, Y. Tanii and M. Tsuda, {\it Phys. Lett.} {\bf B525} (2002) 183. 
%
\bibitem{stt1} K. Shima, Y. Tanii and M. Tsuda, {\it Phys. Lett.} {\bf B546} (2002) 162. 
%
\bibitem{wb} J. Wess and J. Bagger, {\it Supersymmetry and Supergravity} (Princeton University Press, Princeton, New Jersey, 1992). 
%
\bibitem{s2} K. Shima, {\it Phys. Rev.} {\bf D20} (1979) 574. 
%
\bibitem{st4} K. Shima and M. Tsuda, hep-th/0611051. 
%
\bibitem{numass} T. Yanagida, {\it Workshop on Grand Unified Theories}, KEK report 79-18 (1979). 
                  M. Gell-Mann, P. Ramond, R. Slanky, {\it Supergravity} (North Holland, Amsterdam, 1979). 
%
%
%
%
\end{thebibliography}
\end{document}